\documentclass[prl,twocolumn,nofootinbib,superscriptaddress,showpacs,aps]{revtex4-1}% packages

\usepackage{amsmath}
\usepackage{amssymb}
\usepackage{mathtools}
\usepackage{graphicx}
\usepackage{dcolumn,dsfont}% Align table columns on decimal point
\usepackage{bm,braket}% bold math
\usepackage{hyperref}% add hypertext capabilities
\usepackage{isotope}
\usepackage{xspace}% for variable spaces after channel macros
\usepackage{color}
\usepackage{multirow}
\usepackage{tabularx}
\usepackage{subfigure} 
\usepackage{blkarray}
\usepackage{relsize}
\usepackage{bbm}
\usepackage{pbox}
\usepackage{booktabs}
\usepackage{makecell}
\usepackage[export]{adjustbox}
\usepackage{rotating}
\usepackage[activate={true,nocompatibility},final,tracking=true,kerning=true,spacing=true,factor=1100,stretch=10,shrink=10]{microtype}
\usepackage[all]{nowidow}
\hypersetup{
     colorlinks = true,
     linkcolor = cyan,
     citecolor = cyan,
     menucolor = black,
     urlcolor = cyan  
}

\begin{document}

\renewcommand{\vec}[1]{\mathbf{#1}}
\newcommand{\kF}{k_{\rm F}}
\newcommand{\MeV}{\,\mathrm{MeV}}
\newcommand{\keV}{\,\mathrm{keV}}
\newcommand{\GeVi}{\,\mathrm{GeV}^{-1}}
\newcommand{\fm}{\,\mathrm{fm}}
\newcommand{\fmi}{\,\mathrm{fm}^{-1}}
\newcommand{\fmiq}{\,\mathrm{fm}^{-3}}

\newcommand{\etal}{\textit{et~al.}\xspace}
\newcommand{\CC}{C\nolinebreak\hspace{-.05em}\raisebox{.4ex}{\tiny\textbf{+}}\nolinebreak\hspace{-.10em}\raisebox{.4ex}{\tiny\textbf{+}}\xspace}

\title{Chiral interactions up to next-to-next-to-next-to-leading order and nuclear saturation}

\author{C.\ Drischler}
\email[Email:~]{christian.drischler@physik.tu-darmstadt.de}
\affiliation{Institut f\"ur Kernphysik, Technische Universit\"at Darmstadt, 64289 Darmstadt, Germany}
\affiliation{ExtreMe Matter Institute EMMI, GSI Helmholtzzentrum f\"ur Schwerionenforschung GmbH, 64291 Darmstadt, Germany}

\author{K.\ Hebeler}
\email[Email:~]{kai.hebeler@physik.tu-darmstadt.de}
\affiliation{Institut f\"ur Kernphysik, Technische Universit\"at Darmstadt, 64289 Darmstadt, Germany}
\affiliation{ExtreMe Matter Institute EMMI, GSI Helmholtzzentrum f\"ur Schwerionenforschung GmbH, 64291 Darmstadt, Germany}
 
\author{A.\ Schwenk}
\email[Email:~]{schwenk@physik.tu-darmstadt.de}
\affiliation{Institut f\"ur Kernphysik, Technische Universit\"at Darmstadt, 64289 Darmstadt, Germany}
\affiliation{ExtreMe Matter Institute EMMI, GSI Helmholtzzentrum f\"ur Schwerionenforschung GmbH, 64291 Darmstadt, Germany}
\affiliation{Max-Planck-Institut f\"ur Kernphysik, Saupfercheckweg 1, 69117 Heidelberg, Germany}

\begin{abstract}
We present an efficient Monte Carlo framework for perturbative calculations of
infinite nuclear matter based on chiral two-, three-, and four-nucleon
interactions. The method enables the incorporation of all many-body
contributions in a straightforward and transparent way, and makes it possible
to extract systematic uncertainty estimates by performing order-by-order
calculations in the chiral expansion as well as the many-body expansion. The
versatility of this new framework is demonstrated by applying it to chiral
low-momentum interactions, exhibiting a very good many-body convergence up to
fourth order. Following these benchmarks, we explore new chiral interactions
up to next-to-next-to-next-to-leading order (N$^3$LO). Remarkably,
simultaneous fits to the triton and to saturation properties can be achieved,
while all three-nucleon low-energy couplings remain natural. The theoretical
uncertainties of nuclear matter are significantly reduced when going from
next-to-next-to-leading order to N$^3$LO.
\end{abstract}

\maketitle

{\it Introduction.--} Recent calculations of medium-mass and heavy nuclei have
demonstrated the importance of realistic saturation properties of infinite
matter for nuclear forces derived within chiral effective field theory
(EFT)~\cite{Ekst15sat,Hage16NatPhys,Simo16unc,Simo17SatFinNuc,Ekst17deltasat}.
While most nucleon-nucleon (NN) and three-nucleon (3N) interactions fitted to
only two- and few-body observables are able to predict light nuclei in
agreement with experimental data, the theoretical uncertainties tend to
increase with increasing mass number \mbox{$A \gtrsim 16$} (see, e.g.,
Ref.~\cite{Carl15sim}) and significant discrepancies to experiment can be
found for properties of heavy nuclei~\cite{Bind14CCheavy}. There have been
efforts to include properties of heavier nuclei in the optimization of chiral
nuclear forces~\cite{Ekst15sat}. Such interactions tend to exhibit more
realistic saturation properties of nuclear matter and also show improved
agreement with experiment for energies and radii of medium-mass and heavy
nuclei~\cite{Hage16NatPhys,Ruiz16Calcium,Hage16Ni78,Morr17Tin}. However, the
explicit incorporation of nuclear matter properties in the optimization
process of nuclear forces has not been feasible so far due to the
computational complexity of such calculations.

Nuclear matter has been studied based on chiral NN and 3N interactions within
coupled-cluster theory~\cite{Hage14ccnm}, quantum Monte Carlo
methods~\cite{Geze13QMCchi,Rogg14QMC,Lynn16QMC3N}, the self-consistent Green's
function method~\cite{Carb13nm}, and many-body perturbation theory~(MBPT)~\cite{Hebe11fits,Tews13N3LO,Krue13N3LOlong,Holt13PPNP,Cora14nmat,Well14nmtherm,
Dris16asym,Dris16nmatt,Holt16eos3pt}. The advantages of MBPT
are its computational efficiency as well as the possibility to estimate
many-body uncertainties by comparing results at different orders. So far, MBPT
for infinite matter has only been applied up to third order including also the
particle-hole channels~\cite{Cora14nmat,Holt16eos3pt}, where N$^2$LO 3N
contributions beyond Hartree-Fock have been included as normal-ordered
two-body interactions~\cite{Hebe10nmatt,Holt10ddnn,Dris16asym}. Normal
ordering allows to incorporate 3N operators in form of
lower-body operators \cite{Bogn10PPNP}, and nuclear-structure calculations
show that this is an excellent approximation for softer chiral interactions (see, e.g., Refs.~\cite{Hage07CC3N,Roth12NCSMCC3N}). In the MBPT expansion around
Hartree Fock this is a very natural approximation, as the reference state is
sufficiently close to the ground state. There remain
however significant challenges, especially regarding the role of higher-order
particle-hole vs. particle-particle or hole-hole contributions as well as the
inclusion of next-to-next-to-next-to-leading order (N$^3$LO) 3N interactions beyond
Hartree-Fock~\cite{Krue13N3LOlong,Dris16nmatt}.

{\it Novel framework.--} In this Letter, we present a new Monte Carlo
framework for MBPT, which is tailored to address these challenges. We perform
our calculations directly in a single-particle product basis
$\ket{\vec{k}_i\sigma_i \tau_i}$, without needing involved partial-wave
decompositions. Tracing over spin  $\ket{\sigma_i}$ and isospin states
$\ket{\tau_i}$ of each particle with label $i$ is fully automated, whereas the
multidimensional integrals over the momenta $\vec{k}_i$ are computed
using adaptive Monte Carlo
algorithms~\cite{Lepa78Vegas,Hahn05CUBA,Hahn16CUBApara}. This makes
implementing arbitrary energy diagrams straightforward (including
particle-hole contributions), even up to high orders in MBPT, while
approximations in normal ordering are not needed anymore. However, it is well
known that the number of diagrams at each order increases rapidly, with 3, 39,
and 840 at third, fourth, and fifth order for NN-only
interactions~\cite{Stev03autgen,OEISHugen}. Within our Monte Carlo framework,
a manual implementation of these would be feasible but still tedious and at
least inefficient. We therefore developed an automatic \CC code generator based on
the analytic expression of a given diagram.

In addition, we developed a general method to represent chiral interactions
exactly as matrices in spin-isospin space, where the matrix elements are
analytic functions of the single-particle momenta $\vec{k}_i$ in the
programming language~\CC. The automated generation of these
interaction matrices is close to the operatorial definition of chiral
forces~\cite{Mach11PR, Epel02fewbody, Bern083Nlong, Bern113Nshort, Epel064N,Epel074Ndetail,
Epel15NNn4lo,Epel15improved,Ente17EMn4lo},
which we implemented with nonlocal regulators up to N$^3$LO.
For the incorporation of NN interactions
whose operatorial structure is not directly accessible (e.g.,
renormalization-group evolved potentials), we sum the contributions from all
partial-wave channels for each Monte Carlo sampling point.

\begin{figure}[t]
\begin{center}
\includegraphics[page=1,scale=0.96,clip,valign=t]{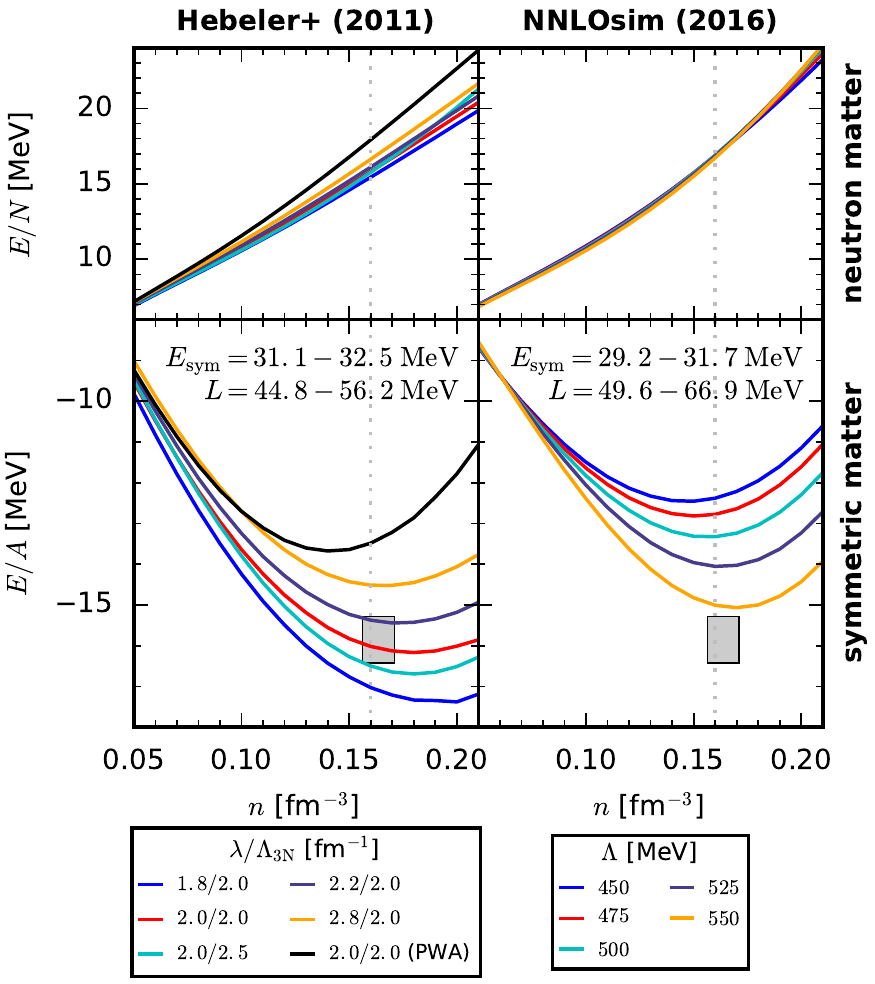}\hspace{5mm}
\end{center}
\caption{Energy per particle of neutron matter (top row) and symmetric nuclear
matter (bottom row) based on the Hebeler+~\cite{Hebe11fits} and NNLOsim~\cite{Carl15sim}
NN and 3N interactions (columns). Results are shown for $\lambda/\Lambda_{\text{3N}}$
for the interactions of Ref.~\cite{Hebe11fits} and $\Lambda = \Lambda_{\text{NN,\,3N}} $ for those of Ref.~\cite{Carl15sim}.}
\label{fig:eos_old}
\end{figure}

\begin{figure}[t]
\begin{center}
\includegraphics[page=1,scale=0.96,clip]{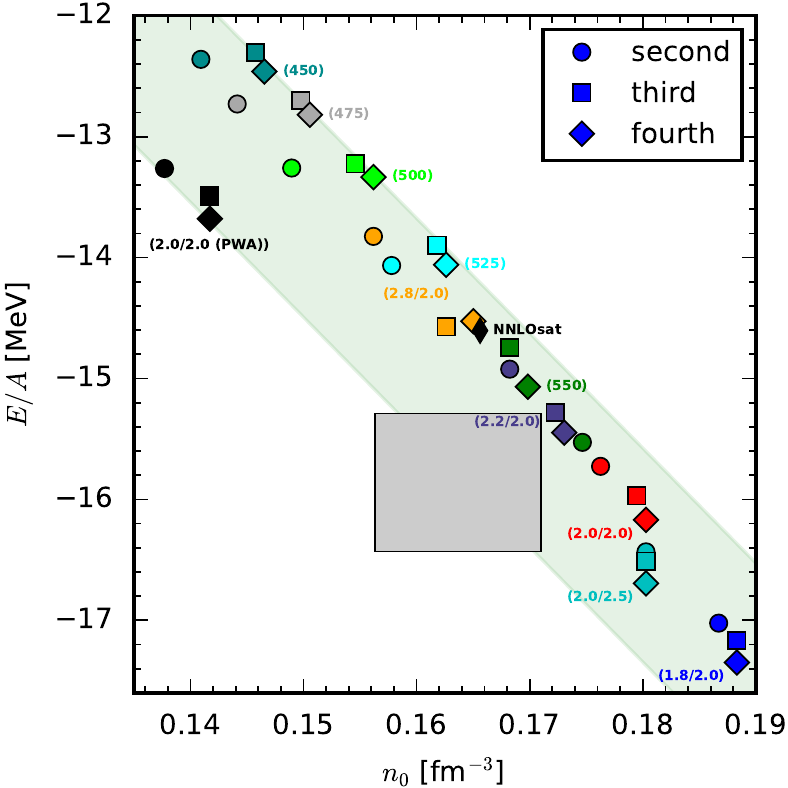}
\end{center}
\caption{Correlation between the calculated saturation density $n_0$ and
saturation energy $E/A$ for the Hebeler+~\cite{Hebe11fits} and NNLOsim~\cite{Carl15sim}
NN and 3N interactions obtained at second, third, and fourth order in MBPT. The
values of $\lambda/\Lambda_{\text{3N}}$ and $\Lambda_{\text{NN}} = \Lambda_{\text{3N}}$,
as well as the saturation region are as in Fig.~\ref{fig:eos_old}. The diamond refers to the
NNLOsat result~\cite{Ekst15sat}.}
\label{fig:coester}
\end{figure}

Specifically, in this first application, we consider all contributions
from NN interactions up to fourth order in MBPT (around the
Hartree-Fock reference state). Contributions from 3N interactions are
included exactly up to second order, including residual 3N-3N terms,
which have only been evaluated so far for contact
interactions~\cite{Kais12exact2nd}. At third order, we neglect all
terms that involve at least one residual 3N contribution, whereas at
fourth order we neglect all 3N contributions. These contributions turn
out to be smaller (see discussion below). This amounts to $4$, $20=3 \times 2^3-4$, and
$24=39-15$ diagrams at second, third, and fourth order, respectively, with up
to 21-dimensional momentum integrals per diagram. 
The number of diagrams at third (fourth) order can be reduced by $4$ ($15$) at zero temperature.
In comparison, a full calculation would involve $39 \times 2^4= 624$ fourth-order
diagrams.

We assess the numerical convergence of the integration by varying the
number of sampling points as well as employing two different
Monte Carlo algorithms~\cite{Hahn05CUBA}, in addition to the variance
as statistical uncertainty. The framework is remarkably efficient due
to performance optimization and parallelization. Most diagrams up to
fourth order can be evaluated within about 10 min to a precision
of $\lesssim 10\keV$. The precise evaluation of a few specific
third-order diagrams involving three 3N interactions requires more
time due to the higher dimensionality of the momentum
integrals. However, the precision can be controlled in a systematic way when
short runtimes are important, e.g., for optimizing nuclear interactions.
Finally, we have performed detailed benchmarks against third-order results
in the literature (see the Supplemental Material~\cite{SuppMat}), including for the dilute Fermi gas~\cite{Hamm00EFT} and
semianalytical as well as partial-wave-based MBPT calculations~\cite{Holt16eos3pt,Dyhd16Regs}.

\begin{table*}[t]
\centering
\caption{Contributions to the energy per particle at $n_0=0.16\fmiq$ in symmetric nuclear matter at
consecutive orders in MBPT based on the Hebeler+~\cite{Hebe11fits} interaction with
$\lambda/\Lambda_\text{3N} = 1.8/2.0~\fmi$ and the N$^2$LO and N$^3$LO interactions of this
work with $\Lambda/c_D$ [for the central $c_D$ fit value (black diamonds) in Fig.~\ref{fig:fits}].
All energies are in MeV.}
\begin{ruledtabular}
\begin{tabular}{ll|ccc|c|cc}
\addlinespace
Chiral order & $\Lambda/c_D$ & \multicolumn{3}{c|}{second order} & \multicolumn{1}{c|}{third order} & 
\multicolumn{2}{c}{fourth order} \\
& & NN-only & NN+3N & 3N res. & NN+3N & NN-only & NN+3N\footnote{Contributions
from 3N forces at fourth order in MBPT are not included in our fits. The values here are an
uncertainty estimate using normal-ordered 3N contributions in the $P=0$ approximation, 
where the center-of-mass momentum of the effective two-body potential is set to
zero~\cite{Hebe10nmatt,Dris16asym}.} \\
\hline
N$^3$LO/N$^2$LO & $\lambda/\Lambda_\text{3N} = 1.8/2.0~\fmi$ & $-2.30$ & $-2.54$ & $-0.10$ & 
$-0.10$ & $-0.20$  & $-0.07$ \\
\cline{1-8}  
\multirow{2}[2]{*}{N$^2$LO} & $450/+2.50$ & $-6.23$ & $-13.38$ & $-0.42$ & $-2.08$ & 
$0.07$  & $0.24$ \\
& $500/-1.50$ & $-8.61$ & $-14.49$ & $-0.66$ & $-0.77$ & $0.32$  & $0.75$ \\
\cline{1-8}    
\multirow{2}[1]{*}{N$^3$LO} & $450/+0.25$ & $-8.84$ & $-14.52$ & $-0.32$ & $-2.28$ & 
$0.61$  & $1.03$ \\ 
& $500/-2.75$ & $-10.56$ & $-14.98$ & $-0.83$ & $-1.05$ & $0.65$  & $1.14$ \\
\end{tabular}%
\end{ruledtabular}%
\label{tab:num_val}%
\end{table*}%

{\it Results for nuclear matter.--} In Fig.~\ref{fig:eos_old} we present results
for the energy per particle in symmetric nuclear matter and neutron matter
based on the Hebeler+~\cite{Hebe11fits} and NNLOsim~\cite{Carl15sim}
NN and 3N interactions up to fourth order in MBPT. For symmetric matter
we show the empirical saturation region by a gray box with boundaries 
$n_0 = 0.164 \pm 0.007\fmiq$ and $E/A = -15.86 \pm 0.37 \pm 0.2\MeV$,
where the first uncertainty is as in Ref.~\cite{Dris16asym} and we add
$0.2\MeV$ from Ref.~\cite{Bert17EstPa}. We also
give results for the symmetry energy $E_\text{sym} =
E/N - E/A$ as well as its slope parameter $L= 3 n_0 \partial_n
E_\text{sym}$ at $n_0=0.16\fmiq$ (dashed vertical line). Both are predicted with narrow ranges.

The Hebeler+ interactions~\cite{Hebe11fits} were obtained by a similarity
renormalization group evolution~\cite{Bogn10PPNP} of the N$^3$LO NN potential
of Ref.~\cite{Ente03EMN3LO} to different resolution scales $\lambda$, whereas
the two leading-order 3N couplings $c_D$ (one-pion-exchange contact
interaction) and $c_E$ (3N contact interaction) were fixed at these resolution
scales by fits to the $^3$H binding energy and the $^4$He charge radius
for two different 3N cutoffs $\Lambda_\text{3N}$. Note that
these potentials include NN (N$^3$LO) and 3N forces (N$^2$LO) up to different
orders in the chiral expansion. Despite being fitted to only few-body data,
these interactions are able to reproduce empirical saturation in
Fig.~\ref{fig:eos_old} within uncertainties given by the spread of the
individual Hebeler+ interactions~\cite{Hebe11fits}. In addition, recent
calculations of medium to heavy nuclei based on some of these
interactions show remarkable agreement with experiment~\cite{Hage16NatPhys,Ruiz16Calcium,
Hage16Ni78,Simo17SatFinNuc,Birk17dipole,Morr17Tin} and thus offer new ab
initio possibilities to investigate the nuclear chart.

The second column of Fig.~\ref{fig:eos_old} shows results for the NNLOsim
potentials~\cite{Carl15sim} ($T_\text{lab}^\text{max}=290\MeV$) for different
cutoff values (see legend). These interactions were obtained by a simultaneous
fit of all low-energy couplings to two-body and few-body data for
$\Lambda_{\text{NN}} = \Lambda_{\text{3N}}$. We observe a weak cutoff
dependence for these potentials in neutron matter over the entire density
range and in symmetric matter up to $n \lesssim 0.08 \fmiq$. At higher
densities, the variation of the energy per particle increases up to $\sim 3
\MeV$ at $n_0=0.16\fmiq$ with a very similar density dependence. Overall, all
the NNLOsim interactions turn out to be too repulsive compared to the
empirical saturation region.

We study the many-body convergence of the Hebeler+ and NNLOsim interactions by
plotting in Fig.~\ref{fig:coester} the calculated saturation energy as a
function of the calculated saturation density at second, third, and fourth
order in MBPT. The annotated values denote the cutoff scales of the different
potentials (see legend of Fig.~\ref{fig:eos_old}). For all low-momentum
interactions as well as with $\Lambda \leqslant 525 \MeV$, we observe a very
good convergence in the many-body expansion, indicating that these chiral
interactions are perturbative over this density regime. Moreover, we find a
pronounced linear correlation similar to the Coester
line~\cite{Coes70nuclmatt}. In contrast to the original Coester line with
NN potentials only, the green band encompassing all (fourth-order) saturation points in Fig.~\ref{fig:coester}
overlaps with the empirical saturation region because of the inclusion of 3N 
forces. Notice, however, that no point lies within the gray box. Note also 
that the Hebeler+ interaction that breaks most from the linear
correlation is ``2.0/2.0 (PWA),'' for which the $c_i$ values in the 3N forces
are larger than in the NN part~\cite{Hebe11fits}.

\begin{figure*}[t]
\begin{center}
\includegraphics[page=1,scale=1.02,clip]{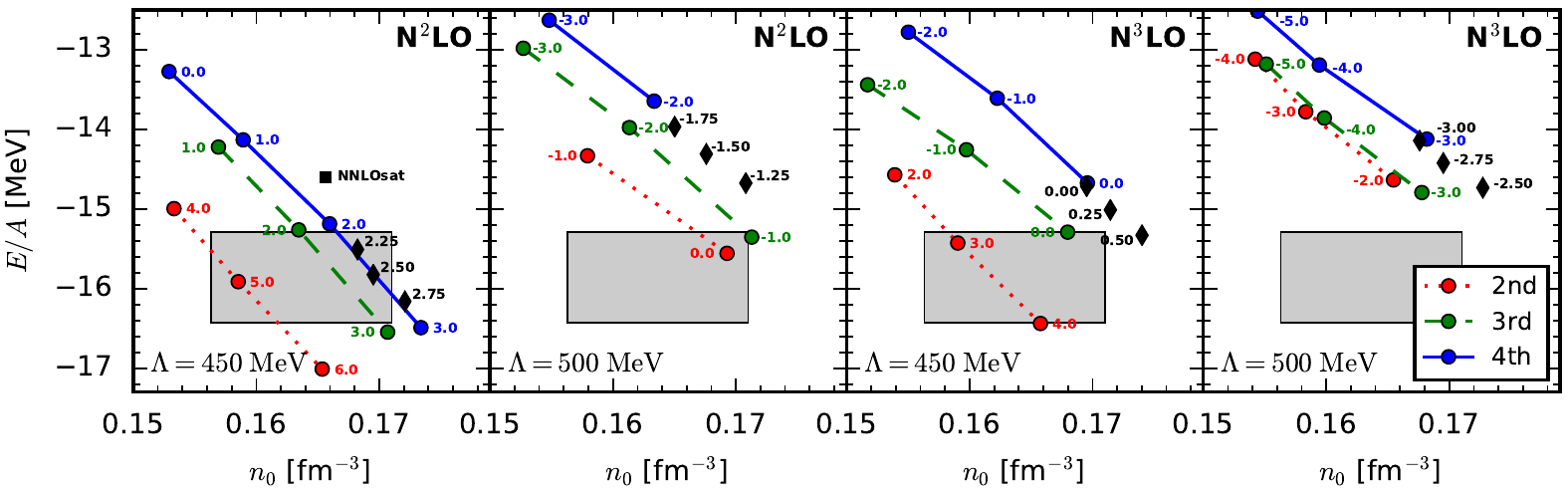}
\end{center}
\caption{Saturation density and energy of symmetric
nuclear matter at different orders in MBPT for the NN and 3N
interactions at N$^2$LO and N$^3$LO. The
points are for different values of $c_D$ (annotated numbers;
$c_E$ follows from Fig.~1 of the Supplemental Material~\cite{SuppMat}), while
the red-dotted, green-dashed, and blue-solid lines correspond to
calculations at second, third, and fourth order in MBPT. The left
(right) two panels are for N$^2$LO (N$^3$LO) with $\Lambda = 450$
and $500 \MeV$. The diamonds in each panel represent the
calculations with a simultaneous good reproduction of both saturation
density and energy at fourth order.}
\label{fig:fits}
\end{figure*}

Finally, in Table~\ref{tab:num_val} we show the hierarchy of contributions
from second, third, and fourth order at $n_0=0.16\fmiq$ for the Hebeler+
``1.8/2.0" interaction, which is most commonly used in the recent \emph{ab initio}
calculations of medium-mass and heavy nuclei. At second order, we give the
contributions from NN interactions (NN-only), from NN plus 3N contributions
that can be represented in form of a density-dependent NN interaction
(NN+3N), and the residual 3N contribution (3N res.). We find that the residual
3N term is significantly smaller compared to the other contributions.
Furthermore, we find that the third-order contributions are significantly
smaller than the second-order terms for all studied interactions. These
findings suggest that the studied interactions exhibit a natural MBPT
convergence pattern for a cutoff of 450~MeV, whereas we already find first
indications of a reduced convergence rate for $\Lambda = 500$~MeV.
Additional higher-order implementations will, however, be necessary to draw final
conclusions on the convergence.

{\it Fit to saturation region.--} The observed convergence pattern indicates
that the studied unevolved nonlocal interactions with $\Lambda \leqslant 525 \MeV$
are sufficiently perturbative and allow calculations with controlled many-body
uncertainties. This offers the possibility to use the new Monte Carlo
framework for constraining the 3N couplings using information from nuclear
matter. In this Letter, we demonstrate this using the N$^2$LO and N$^3$LO NN
potentials of Entem, Machleidt, and Nosyk (ENM)~\cite{Ente17EMn4lo} with
$\Lambda_\text{NN} = 450$ and $500\MeV$, which are also very promising in
terms of their Weinberg eigenvalues~\cite{Hopp17WeinEVAn}. As a first step, we
fit to the $^3$H binding energy that leads to a relation of the 3N couplings
$c_D$ and $c_E$ (shown in Fig.~1 of the Supplemental Material~\cite{SuppMat}). For the fits, we include all 3N
contributions consistently up to N$^2$LO and N$^3$LO, respectively. The
corresponding 3N matrix elements were computed as in Ref.~\cite{Hebe15N3LOpw}.
We use $\Lambda = \Lambda_{\text{NN,\,3N}}$ and a nonlocal
regulator of the form $f_{\Lambda} (p,\,q) = \exp [ - ( (p^2 + 3/4
q^2)/\Lambda_\text{3N}^2)^4 ]$ for the Jacobi momenta $p$ and $q$ of the initial and
final states~\cite{Epel02fewbody}. For both cutoffs and chiral orders, we
obtain $c_E$ couplings of natural size in the wide $c_D$ range explored.

As a second step, we calculate nuclear matter for the range of 3N couplings
and determine the saturation point. In Fig.~\ref{fig:fits}, we present the
saturation points at N$^2$LO and N$^3$LO as a function of $c_D$ and at
different orders in MBPT. Similar to the interactions shown in
Fig.~\ref{fig:coester}, we find a natural convergence pattern. Note that the
shown points on the trajectories correspond to different $c_D$ values at
second order compared to third and fourth order. Contributions at third order
are therefore more significant in these cases, whereas fourth-order
corrections are again much smaller as is shown in Table~\ref{tab:num_val}
(the convergence at fixed densities is documented in Table~I of
the Supplemental Material~\cite{SuppMat}). In general, Fig.~\ref{fig:fits} demonstrates that it
is possible to determine natural $c_D$/$c_E$ combinations at N$^2$LO and
N$^3$LO with good saturation properties for both cutoff cases considered.
However, N$^3$LO contributions provide
slightly too much repulsion.

\begin{figure}[ht!]
\begin{center}
\includegraphics[page=1,scale=0.96,clip]{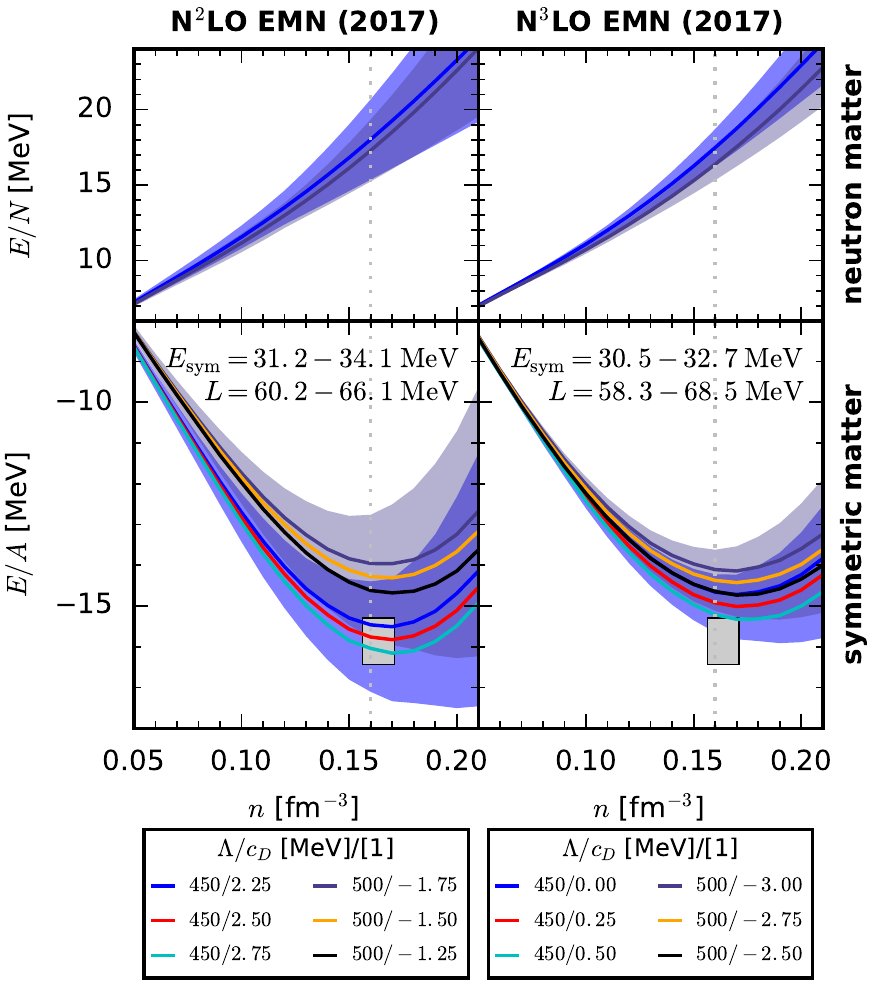}
\end{center}
\caption{Energy per particle in neutron matter (top row) and
symmetric nuclear matter (bottom row) based on chiral interactions at N$^2$LO
(first column) and N$^3$LO (second column) fit to the empirical saturation
region (see Fig.~\ref{fig:fits}). The fits are labeled by $\Lambda/c_D$ in the
legend. The blue
($\Lambda=450~\MeV$) and gray ($\Lambda=500~\MeV$) bands estimate the
theoretical uncertainty following Ref.~\cite{Epel15improved}. 
Note that the annotated results for $E_\text{sym}$ and $L$ do not include this uncertainty.}
\label{fig:eos_new}
\end{figure}

In each panel of Fig.~\ref{fig:fits}, we mark the three couplings that provide
a good fit to the saturation region by black diamonds, with annotated $c_D$
values (the corresponding $c_E$ values are given in Fig.~1 of the Supplemental Material~\cite{SuppMat}). The
resulting equations of state of symmetric nuclear matter and neutron matter at
N$^2$LO and N$^3$LO are shown in Fig.~\ref{fig:eos_new}. Note that only two
lines are present in neutron matter since the shorter-range 3N interactions do
not contribute~\cite{Hebe10nmatt}. We also calculate the
Hartree-Fock energy of the N$^3$LO 4N forces using the nonlocal regulator as in
Ref.~\cite{Krue13N3LOlong}. These forces are long range and free of unknown
parameters~\cite{Epel064N,Epel074Ndetail}. The obtained 4N Hartree-Fock
energies at $n_0$ are $\approx -(150-200)\keV$ in neutron matter as well as
symmetric matter, in agreement with the results of Ref.~\cite{Krue13N3LOlong}.
As for the Hebeler+ and NNLOsim results, the symmetry energy and the $L$
parameter are predicted with a remarkably narrow range. In symmetric matter,
we also observe a weak cutoff dependence at N$^3$LO, whereas the results for
$\Lambda=450\MeV$ are clearly separated from $\Lambda=500\MeV$ at N$^2$LO,
with the former achieving the best fits to the saturation region. Finally, we
estimate the theoretical uncertainty from the chiral expansion following
Ref.~\cite{Epel15improved}, using $Q = p/\Lambda_b$ with breakdown scale
$\Lambda_b = 500 \MeV$ and average momentum $p =\sqrt{3/5} \, \kF$. The bands
overlap from N$^2$LO to N$^3$LO, and we clearly see that the uncertainties are
significantly reduced at N$^3$LO. For reference, results at LO and NLO are shown in
Fig.~2 of the Supplemental Material~\cite{SuppMat}.

{\it Summary.--} We have presented a new Monte Carlo framework for
calculations of nuclear matter, which allows to include higher-order
contributions from chiral interactions and is capable of going to high
enough orders in the many-body expansion for suitable
interactions. The new method was applied to the calculation of the
symmetric-matter and neutron-matter energy in an expansion around
Hartree-Fock, but it can be easily generalized to expansions around
other reference states. This enabled first benchmarks of chiral
low-momentum interactions to fourth order in MBPT showing a systematic
order-by-order convergence. We then used this to develop new chiral
interactions at N$^2$LO and N$^3$LO, including NN, 3N, and 4N
interactions at N$^3$LO, where the 3N couplings are fit to the triton
and to saturation properties. Our work shows that a good description
of nuclear matter at these orders is possible, with a systematic
behavior from N$^2$LO to N$^3$LO and natural low-energy couplings.
Thanks to the computational efficiency, the new framework is also
ideal for the incorporation of nuclear matter properties in the
fitting of novel nuclear interactions. It will be exciting to see
what these interactions predict for nuclei and for the equation of
state for astrophysics.

\begin{acknowledgments}

We thank A.~Ekstr{\"o}m, B.~Carlsson, C.~Forss{\'e}n, R.~J.~Furnstahl, and 
T.~Hahn for useful discussions, R.~Machleidt for providing us with the
EMN potentials, and J.~W.~Holt for benchmark values at third order in
MBPT. We also thank C.~Iwainsky for helping us to optimize the code
performance. This work was supported in part by the European Research
Council Grant No.~307986 STRONGINT and the Deutsche
Forschungsgemeinschaft through Grant SFB~1245. Computational resources
have been provided by the Lichtenberg high performance computer of the
TU Darmstadt.

\end{acknowledgments}

\bibliographystyle{apsrev4-1}

\bibliography{strongint}

\end{document}